\documentclass[aps,prb,twocolumn,showpacs,a4paper,superscriptaddress]{revtex4-2}
\usepackage{graphicx,bm,amsmath,dcolumn,amssymb}
\usepackage{longtable}
\graphicspath{{figures}}
\usepackage{geometry,hyperref,color}
\begin{document}
\title{Dynamical relaxation of a long-range XY chain}
  \author{Yu-Huan Huang}
  \affiliation{School of Microelectronics $\&$ Data Science, Anhui University of Technology, Maanshan 243002, China }
  \author{Yin-Tao Zou}
  \affiliation{School of Microelectronics $\&$ Data Science, Anhui University of Technology, Maanshan 243002, China }
   \author{Chengxiang Ding}
  \email{dingcx@ahut.edu.cn}
   \affiliation{School of Microelectronics $\&$ Data Science, Anhui University of Technology, Maanshan 243002, China }
   \affiliation{Anhui Provincial Joint Key Laboratory of Disciplines for Industrial Big Data Analysis and Intelligent Decision, Maanshan 243002, China}
\date{\today}
\begin{abstract} 
  We study the universal real-time relaxation behaviors  of a  long-range quantum XY chain following a  quench.
  Our research includes both the noncritical and  critical quench. 
  In the case of noncritical quench, i.e., neither the initial state nor the postquench Hamiltonian is at a critical point of equilibrium phase transition,
  a quench to the commensurate phase or incommensurate phase gives a  scaling of $t^{-3/2}$ or $t^{-1/2}$, respectively, 
  which is the same as the counterpart of the short-range XY model.
 However, for a quench to the boundary line between the commensurate and incommensurate phases, 
  the scaling law $t^{-\mu}$ may be different from the  $t^{-3/4}$ law of the counterpart of the short-range model.
   More interestingly, the decaying exponent $\mu$ may depend on the choice of the parameters of the postquench Hamiltonian because of the different  asymptotic behaviors of the energy spectrum. 
   Furthermore, in certain cases, the scaling behavior may be outside the range of predictions made by the stationary phase approximation, because an inflection point emerges in the energy spectrum.
  For the critical quench, i.e., the initial state or the postquench Hamiltonian is at a critical point of equilibrium phase transition, the aforementioned scaling law $t^{-\mu}$
  may be changed because of the gap-closing property of the energy spectrum of the critical point.

\end{abstract}
\maketitle 

\section{Introduction}
In the study of equilibrium state phase transitions, it is well-known that the critical exponents of a phase transition are only determined by  the symmetries of the order parameter and the space dimension, but not depend on the details of the system, this is called ``universality" , which is a very  important conception in the study of phase transition and critical phenomena.   More and more researches show that such conception is also increasingly showing its importance in the study of nonequilibrium physics, such as the first\cite{DQPT-I,DQPT-Ia} and second\cite{Heyl2013,Heyl2015,ding2020,DQPT-IIx} types of dynamical quantum phase transitions, Kibble-Zurek mechanism\cite{Kibble1976,Kibble1980,Zurek1985,Zurek1996,KZ}, phase ordering\cite{PO,PO1,PO2,Janke2019,Janke2020}, and relaxation processes\cite{DQPT-3a,pbc,apbc,stoch,noise,quench1,quench2}.
In the study of these phenomena, the dynamical behavior of some physical quantities may exhibit power-law scaling similar to the equilibrium critical phenomena, which can also define universal indexes like critical exponents.

Taking phase ordering as an example\cite{PO1,PO2,Janke2019},  the characteristic length $l(t)$ of the short-range Ising model scales as $l(t) \sim t^b$ after a sudden quench from a high-temperature disordered phase  to a low-temperature ordered phase,  where the growth index $b=1/2$;  for the conserved Ising, this index is $b$=1/3.  The index $b$ is universal, independent of the details of the model, and does not depend on the spatial dimension of the model; but for long-range models, the index is different\cite{PO1}. 

In the quench of  quantum isolated systems, the situation is different\cite{quench1,quench2}. Since unitary evolution does not destroy the symmetry of the model, quantum quench cannot lead to a phase ordering process similar to classical models; however, in the process of quantum relaxation approaching steady state, the difference between the instantaneous and steady-state values of some physical quantities may also follow a  power-law decaying behavior, and the decaying index is also universal, which does not depend on the details of the model, but mainly depends on the  structure of the  energy spectrum of the postquench Hamiltonian. For example,  in the study of the relaxation process of the quantum XY chain\cite{quench1}, Makki and his collaborators point out that the decaying index of the short-range correlation can be obtained through stationary phase approximation (SPA), with the decaying indices of -3/2 and -1/2 for the quenches to commensurate  and incommensurate phases, respectively. Similar relaxation processes have also been studied in quantum driven systems, such as the periodic and antiperiodic driven systems\cite{pbc,apbc}, as well as the random and noise driven systems\cite{stoch,noise}. In our recent research\cite{Ding2023}, we found that a quantum quench from a critical ground state may lead to changes in the aforementioned critical indexes due to the closing of the energy gap at  the critical point; at the same time, we also discovered interesting crossover behaviors related to such type of quench.

 In this paper, we extend our research to  the long-range integrable systems, paying special attention to  the influence of long-range interactions on the universal relaxation  behaviors. We find that under certain situations, long-range interactions may lead to very different structures of the energy spectrum compared to the short-range model, 
 correspondingly, the universal relaxation behaviors  are very different.
 Especially, the long-range interactions can  lead to inflection point in the energy spectrum, which makes  the relaxation behavior go beyond the prediction of the  stationary phase approximation. Our research includes both the noncritical and  critical quench.

The paper is arranged as follows: In Sec. \ref{models} and Sec. \ref{method}, we introduce the long-range XY model we studied and the method we adopt, respectively; in Sec. \ref{noncri} and \ref{cri}, we give the results of the noncritical quench  and critical quench, respectively; we conclude our paper in Sec. \ref{con}.
\section{Models}
\label{models}
The models we studied is a long-range XY model, whose Hamiltonian is written as
\begin{eqnarray}
	H&=&-\frac{1}{\mathcal{N}}\sum\limits_{j=1}^L\sum\limits_{m=1}^MJ_m(\frac{1+\chi}{2}\sigma_j^x\sigma_{j+m}^x+\frac{1-\chi}{2}\sigma_j^y\sigma_{j+m}^y)\nonumber\\
	&&\times O^z_{j+1,m-1}-h\sum\limits_{j=1}^N\sigma_j^z,\label{longXY}
\end{eqnarray}
where $\sigma^x,\sigma^y$, and $\sigma^z$ are the Pauli matrixes, and $L$ the number of sites of the chain, in which the periodic boundary condition is applied.  $M=L/2-1$, the operator $O^z_{j+1,m-1}=\prod_{n=j+1}^{j+m-1}\sigma_n^z$, and the coupling $J_m=Jm^{-\alpha}$, where we set $J=1$. In current paper, we only consider the case of $\alpha>1$. The normalization constant $\mathcal{N}$ is
\begin{eqnarray}
	\mathcal{N}=\sum\limits_{m=1}^{M}J_m,
\end{eqnarray}
this  preserves the extensivity of the Hamiltonian.
 
By the Jordan-Winger transformation, the model can be transformed to a free-fermion model
\begin{eqnarray}
	H_f&=&-\frac{1}{\mathcal{N}}\sum\limits_{j=1}^L\sum\limits_{m=1}^MJ_m(c_j^\dagger c_{j+m}+\chi c_j^\dagger c_{j+m}^\dagger+{\rm H.c.})\nonumber\\
	 &&+h\sum\limits_{j=1}^L(2c_j^\dagger c_j-1),\label{longfermion}
\end{eqnarray}
here, we have restricted our study in the even-fermionic-number-parity sector and adopted the antiperiodic boundary condition.

By the Fourier transiformation, the model can be transformed to the momentum space, 
\begin{eqnarray}
	H=\sum\limits_{k>0}H_k=\sum_{k>0}\mathbf{\Psi}_k^\dagger \mathbf{H}_k\mathbf{\Psi}_k
\end{eqnarray}
where  $\mathbf{\Psi}_k=(c_k,c^\dagger_{-k})^T$ are Nambu spinors and 
\begin{eqnarray}
	\mathbf{H}_k=
	\begin{pmatrix}
		z_k & -iy_k\\
		iy_k & -z_k
	\end{pmatrix}.
\end{eqnarray}
The wave vector $k$ belongs to $\{  \pm(2n-1)\pi/L, n=1, 2, \cdots, L/2 \}$ and 
\begin{eqnarray}
z_k&=&2\big[h-\frac{1}{\mathcal{N}}\cdot\sum_{m=1}^MJ_m\cos(mk)\big],\\
y_k&=&2\frac{\chi}{\mathcal{N}}\cdot\sum_{m=1}^MJ_m\sin(mk)). 
\end{eqnarray}

The hamiltonian $H_k$ is already in a small Hilbert space of $2\times 2$, it can be easily diagonalized by the Bogoliubov transformation 
\begin{eqnarray}
\gamma_k&=&u^*_kc_k+v^*_kc_{-k}^\dagger,\\
\gamma_{-k}^\dagger&=&-v_kc_k+u_kc_{-k}^\dagger,
\end{eqnarray}
where $u_k=\cos(\theta_k/2), v_k=i\sin(\theta_k/2)$, with $\theta_k$ the Bogoliubov angle defined as 
\begin{eqnarray}
\tan\theta_k=y_k/z_k \label{theta}
\end{eqnarray}
This gives the energy spectrum and the ground state of the model
\begin{eqnarray}
	\varepsilon_k&=&\sqrt{y_k^2+z_k^2},\\
	|\Phi\rangle&=&\prod\limits_{k>0}(u_k+v_kc_k^\dagger c_{-k}^\dagger )|0\rangle.
	\end{eqnarray}

 \begin{figure}[thpb]
	\centering
	\includegraphics[width=1\columnwidth]{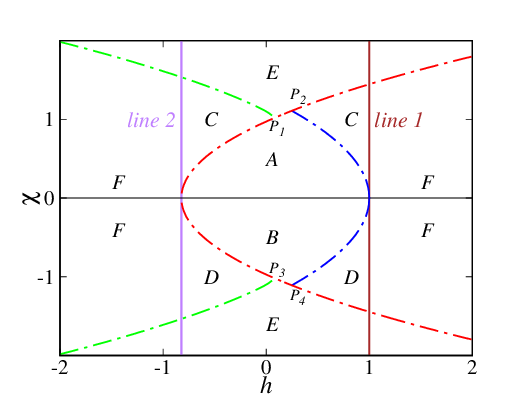}
	\caption{(Color online) Equilibrium phase diagram of model (\ref{longXY}) with $\alpha=3.5$:  the solid lines are critical lines between the ordered and disordered phases; A, B, and E are incommensurate phase, C, D, and F are commensurate phase; the dashed lines are the boundaries between the incommensurate and commensurate phases. }
	\label{pd}
\end{figure}
\begin{figure}[htpb]
	\centering
	\includegraphics[width=1.0\columnwidth]{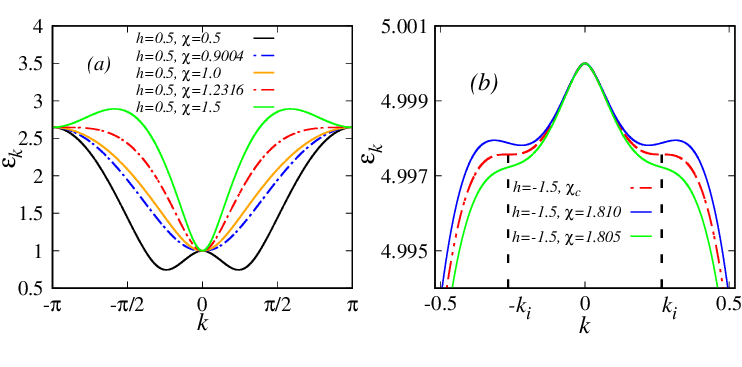}
	\caption{(a) Energy spectrum of several Hamiltonians of the long-range XY model (\ref{longXY}) at $h=0.5$, with $\alpha=3.5$; (b) Energy spectrum of three Hamiltonians with $h=-1.5$, where $\chi_c=1.8076$ and $k_i=0.26625$.} 
	\label{spec0}
\end{figure}

 
The equilibrium phase diagram of the two models are shown in Fig.  \ref{pd}, the  solid line 1 and line 2 are the two critical lines determined  by the gap-closing momentum  $k=0$ and $\pi$, respectively, which are written as
\begin{eqnarray}
	&&h_{\rm c}^{(1)}=1,\\
	&&h_{\rm c}^{(2)}=\lim\limits_{L\rightarrow\infty}\frac{1}{\mathcal{N}}\cdot\sum_{m=1}^{L/2-1}J_m(-1)^k.
\end{eqnarray}
The dashed lines are the boundaries between commensurate phase and the incomensurate phase, where the  incommensurate phase is defined if there is an additional saddle point $k_0$ in the energy spectrum besides the saddle points $k=0$ and $\pi$, otherwise, it is a commensurate phase; 
the boundary is where the saddle point $k_0$ just disappears; this is demonstrated in Fig. \ref{spec0}(a).  The red and blue dashed lines are the boundaries 
where the saddle point $k_0$ merges to $k=\pi$ and $0$, respectively.  However, the green dashed line is very special,  on this line, the additional saddle point $k_0$ also disappears but in a different way, which is demonstrated in Fig. \ref{spec0}(b); we can see that in this case, along with the disappearance of the saddle point $k_0$, an inflection point $k_i$ appears. As we will show later, such inflection point leads to very substantial change in the scaling behavior of relaxation. 

\section{Method}
\label{method}
The evolution of a quantum state $|\Phi(t)\rangle$ 
is determined by the  time-dependent Schr\"{o}dinger equation, for the quantum  XY model,  this is equivalent to solving the time-dependent  Bogoliubov–de Gennes (BdG) equation\cite{begin}.
\begin{eqnarray}
	i\partial_t\phi_k(t)=\mathbf{H}_k\phi_k(t)
\end{eqnarray}
where $\phi_k(t)=(v_k(t), u_k(t)  )^T$. For the question of a sudden quench, 
the solution is a unitary evolution
\begin{eqnarray}
	\phi_k(t)=e^{-i\mathbf{H}_kt}\phi_k(0),\label{uniev}
\end{eqnarray}
where $\phi_k(0)=(v_k(0), u_k(0))^T$.   After some simple algebraic calculations, we get
\begin{eqnarray}
	v_k(t)&=&v_k(0)[\cos(\varepsilon_kt)-i\sin(\varepsilon_kt)\cos\theta_k]\nonumber\\
	&&-u_k(0)\sin(\varepsilon_kt)\sin\theta_k\\
	u_k(t)&=&u_k(0)[\cos(\varepsilon_kt)+i\sin(\varepsilon_kt)\cos\theta_k]\nonumber\\
	&&+v_k(0)\sin(\varepsilon_kt)\sin\theta_k.
\end{eqnarray}
Using the expression of $v_k(t)$, the correlation  $C_{mn}(t)=\langle c_m^\dagger c_n\rangle$ can be calculated, which is 
\begin{eqnarray}
	C_{mn}(t)&=&\frac{1}{\pi}\int_{0}^{\pi}dk|v_k(t)|^2\cos[k(m-n)]\nonumber\\
	&=&C_{mn}(\infty)+\delta C_{mn}(t),
\end{eqnarray}
where 
\begin{eqnarray}
	C_{mn}(\infty)=&&\frac{1}{\pi}\int_{0}^{\pi}dk\Big[1-\cos\zeta_k\cos^2\theta_k\nonumber\\
	&&-\frac{1}{2}\sin\zeta_k\sin(2\theta_k)\Big]\cos[k(m-n)]
\end{eqnarray}
is the value  of $C_{mn}(t)$ in the  steady state. Here $\zeta_k$ is the Bogoliubov angle of the prequench Hamiltonian, $\theta_k$ is the Bogoliubov angle of the postquench Hamiltonian.  $\delta C_{mn}(t)$ is the difference  between $C_{mn}(t)$ and $C_{mn}(\infty)$,  
\begin{eqnarray}
	\delta C_{mn}(t)=\delta C_{mn}^{(1)}(t)+\delta C_{mn}^{(2)}(t),
\end{eqnarray}
where 
\begin{flalign}
	&	\delta C_{mn}^{(1)}(t)\nonumber\\
	&  =-\frac{1}{\pi}\int_{0}^{\pi}dk\cos\zeta_k\sin^2\theta_k\cos(2\varepsilon_kt)\cos[k(m-n)]\nonumber\\
	& ={\rm Re}\Big\{-\frac{1}{\pi}\int_{0}^{\pi}dkF^{(1)}_ke^{2i\varepsilon_kt}\cos[k(m-n)]\Big\}, \label{int1}
\end{flalign}
with 
\begin{eqnarray}
 F^{(1)}_k=\cos\zeta_k\sin^2\theta_k;
\end{eqnarray}
and 
\begin{flalign}
	&	\delta C_{mn}^{(2)}(t)\nonumber\\
	& =\frac{1}{2\pi}\int_{0}^{\pi}dk\sin\zeta_k\sin(2\theta_k)\cos(2\varepsilon_kt)\cos[k(m-n)]\nonumber\\
	& ={\rm Re}\Big\{\frac{1}{2\pi}\int_{0}^{\pi}dkF^{(2)}_ke^{2i\varepsilon_kt}\cos[k(m-n)]\Big\}. \label{int2}
\end{flalign}
with 
\begin{eqnarray}
	F^{(2)}_k=\sin\zeta_k\sin(2\theta_k).
\end{eqnarray}

The asymptotic behavior of $|\delta C_{mn}(t)|$ is the main topic of the current paper, which can be obtained by the SPA\cite{quench1},  where the key point is that the integral in $\delta C_{mn}^{(1)}(t)$ or $\delta C_{mn}^{(2)}(t)$ is dominated by the contributions near the extrema of $\varepsilon_k$, and the factor $e^{2i\varepsilon_kt}$ is replaced by a Gaussian by the Taylor expansion of $\varepsilon_k$ at the extrema $k_0$,  where $k_0$ in general is a saddle point. Then the integrals are calculable, and the scaling behaviors can be obtained.  
In summary, when $t$ is large enough, the integral (\ref{int1}) or (\ref{int2}) is approximately proportional to 
\begin{eqnarray}
	\small
	{\rm Re}\Bigg\{e^{i(2\varepsilon_{k_0}t+\varphi)}\int_{-\infty}^\infty dk (k-k_0)^q \exp\big[ib (k-k_0)^pt\big]\Bigg\}, \label{int3}
\end{eqnarray}
where  $\varphi$ and $b$ are trivial constants; $p$ is determined by the asymptotic behaviors  of the spectrum of the postquench Hamiltonian in the vicinity of the saddle point, i.e., 
\begin{eqnarray}
	\varepsilon_k\sim \varepsilon_{k_0}+b(k-k_0)^p,
\end{eqnarray}
$q$ is determined by the asymptotic behaviors of the factors $F^{(1)}_k$ and $F^{(2)}_k$
 as $k$ approaches $k_0$, i.e., 
 \begin{eqnarray}
 	&&F_k^{(1)}\sim (k-k_0)^q,\\
    &&F_k^{(2)}\sim (k-k_0)^q.
 \end{eqnarray}
    Then the scaling behavior of $|\delta C_{mn}(t)|$ in the long-time limit is 
\begin{eqnarray}
	&&|\delta C_{mn}(t)|\sim t^{-\mu},\\
	&&{\rm with} ~ \mu=(q+1)/p.
\end{eqnarray}
For example, for a  noncritical quench to the commensurate phase, in the vicinity of the saddle point $k=0$, 
$p=q=2$,  which eventually leads to a power law of $t^{-3/2}$.
For more details,  see the Appendix A of Ref. \onlinecite{quench1}, where more examples are given.  
However, in Ref. \onlinecite{quench1} the analysis is restricted to $\delta C_{mn}^{(1)}(t)$, because the initial state is chosen as $(v_k(0),u_k(0))$=$(0,1)$.  
Generally, $|\delta C_{mn}^{(2)}(t)|$ follows the same scaling law of $|\delta C_{mn}^{(1)}(t)|$, however, for the critical quench\cite{Ding2023}, 
the scaling laws can be different, because the gap-closing property of $\varepsilon_k$ may substantially change the asymptotic  behavior of the factor $F^{(1)}_k$ or $F^{(2)}_k$.  In current paper,  we will show that the long-range interactions can also change the asymptotic behaviors of the two factors and also the asymptotic  behavior of the energy spectrum in certain cases, which lead to new scaling behaviors in the relaxation.

In the current paper,  in the calculations of  integrals (\ref{int1}) and (\ref{int2}), $m$ is set to be equal to $n$  if not explicitly stated.  The results do not have qualitative difference for $m\ne n$ if the distance between the sites $m$ and $n$ is short. 

\section{Noncritical quench}
\label{noncri}
\subsection{Quench to incomensurate and commentsurate phases}
\begin{figure}[htpb]
 	\centering
 	\includegraphics[width=1\columnwidth]{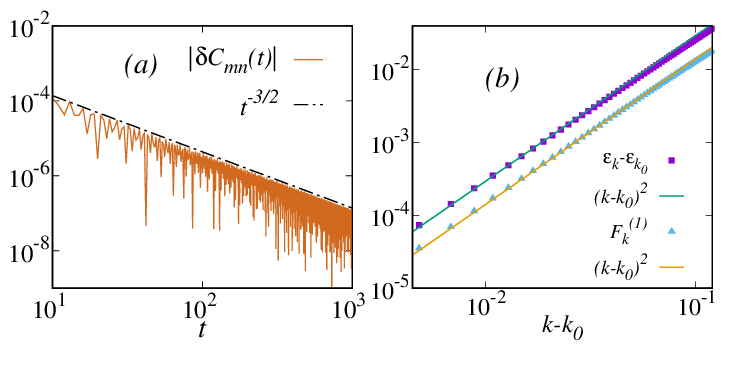}
 	\caption {(a) Scaling behavior of quench from $(h, \chi)$=(3, 1) to (1.5, 0.5), where $\alpha=3.5$ and $L=20000$; (b) Asymptotic behaviors of the energy spectrum and the factor $F_k^{(1)}$ in the vicinity of  the saddle point $k_0=0$.}
 	\label{ncq1a}
 \end{figure}

Before investigating the dynamical relaxation behaviors of the critical quench of the long-range XY model (\ref{longXY}), we study the noncritical quench of this model at first. The $t^{-3/2}$ and $t^{-1/2}$ scaling laws are found for the quench to the commensurate phase and incommensurate phase, respectively. 
For example, as shown in Fig. \ref{ncq1a}(a),  for a quench from $(h,\chi)$=(3,1) to (1.5,0.5) with $\alpha=3.5$, $|\delta C_{mn}(t)|$ satisfies 
\begin{eqnarray}
|\delta C_{mn}(t)|\sim t^{-3/2}.
\end{eqnarray}
This result falls in the prediction of SPA. As shown in Fig. \ref{ncq1a}(b), in the vicinity of saddle point $k_0=0$, the asymptotic behavior of the energy spectrum  and $F_{k}^{(1)}$  satisfies 
\begin{eqnarray}
\varepsilon_k-\varepsilon_{k_0}\propto(k-k_0)^2,\\
F_{k}^{(1)}\propto (k-k_0)^2,
\end{eqnarray}
this gives $p=q=2$ for Eq. (\ref{int3}),  subsequently the scaling of $|\delta C_{mn}(t)|$  is $t^{-3/2}$.

Similarly, we investigate the quench to the incommensurate phase; for example, for a quench  from $(h,\chi)$=(3,1) to (0.5,0.5) with $\alpha=3.5$, 
the scaling behavior is
 \begin{eqnarray}
 	|\delta C_{mn}(t)|\sim t^{-1/2},
 \end{eqnarray}
in this case, the saddle point that governs the scaling behavior is $k_0=0.2435\pi$, which is not equal to 0 or $\pi$; therefore, the asymptotic behaviors of the energy spectrum and  $F_{k}^{(1)}$ are 
\begin{eqnarray}
	&&\varepsilon_k-\varepsilon_{k_0}\propto(k-k_0)^2,\\
	&&F_{k}^{(1)} \sim {\rm constant},
\end{eqnarray}
this gives $p=2$ and $q=0$ for Eq. (\ref{int3}),  subsequently the scaling of $|\delta C_{mn}(t)|$  is $t^{-1/2}$.

These results  are similar to the corresponding cases of the short-range XY model. 
 More importantly,  here the scaling exponents do not vary with the change of $\alpha$ (with $\alpha>1$). The only effect of the smaller value of $\alpha$ is that the revival phenomena\cite{revival} may appear at earlier time.

\subsection{Quench to the dashed boundary lines}
For a quench to the dashed boundary lines of Fig. \ref{pd},  we find that the scaling of $|\delta C_{mn}(t)|$ depends on the choosing of the parameters of the postquench Hamiltonian.
For example,  for a quench from $(h,\chi)$=(3, 1) to (0.5, 1.2316), in which the postquench Hamiltonian is on the red dashed line,
the scaling is 
\begin{eqnarray}
|\delta_{mn}(t)|\sim t^{-3/4}, \label{ncred}
\end{eqnarray}
which is the same as  corresponding case of short-range XY model.
In this case,   it is easy to verify that the asymptotic behaviors of the energy spectrum and the  factor $F_k^{(1)}$ satisfy
\begin{eqnarray}
&&\varepsilon_k-\varepsilon_{k_0}\propto(k-k_0)^4, \label{rdb1}\\
&&F_k^{(1)}\propto (k-k_0)^2,\label{rdb2}
\end{eqnarray}
where $k_0=\pi$; therefore,   we get  $q=2$ and $p=4$ for the integral (\ref{int3}), subsequently the scaling is $t^{-3/4}$.

\begin{figure}[htpb]
	\centering
	\includegraphics[width=0.75\columnwidth]{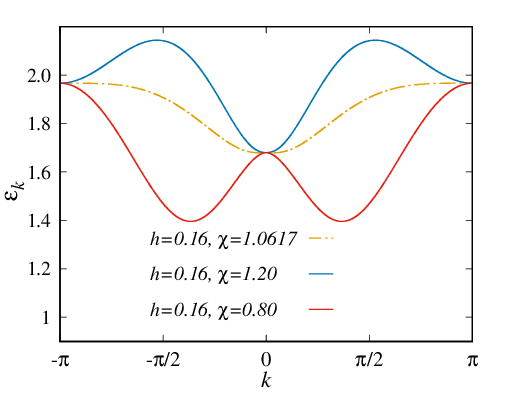}
	\caption{Energy spectrums of three Hamiltonians, the dashed line is the spectrum of a Hamiltonian with ($h,\chi$)=(0.16, 1.0617), which is a point on the curve segment $p_1p_2$.} 
	\label{specA}
\end{figure}

We further check several cases with the postquench Hamiltonian on the red dashed line  of Fig. \ref{pd},  it is shown that the $t^{-3/4}$ scaling always keeps,
this includes the cases on the  curve segment $p_1p_2$ and $p_3p_4$. 
It should be noted that the two curve sections are very special, because both the two sides of the curve sections are incommensurate phases; 
for an  intuitive understanding of this results,  we plot the three energy spectrums of the Hamiltonians near the curve section $p_1p_2$ , as  shown in Fig.  \ref{specA}.
In the vicinity of the saddle point $k=\pi$, the asymptotic behaviors of the energy spectrum and the factor $F_k^{(1)}$ of the points on the curve section $p_1p_2$ and $p_3p_4$ also satisfy Eqs. (\ref{rdb1}) and (\ref{rdb2}), respectively, therefore, we conclude that the scaling of the whole red dashed line is $t^{-3/4}$.

\begin{figure}[htpb]
	\centering
	\includegraphics[width=1.0\columnwidth]{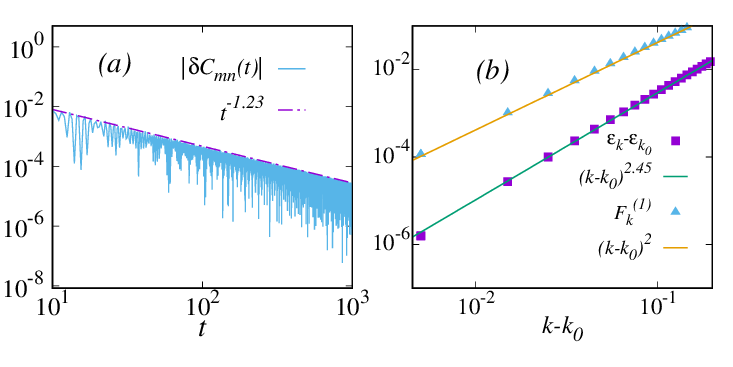}
	\caption{ (a) Scaling behavior of a quench from $(h,\chi)$=(3,1) to (0.5,0.90041), where the postquench Hamiltonian is a point on the blue dashed line of Fig. \ref{pd}; (b) Asymptotic behaviors of the energy spectrum of the postquench Hamiltonian and the factor $F_k^{(1)}$, where $k_0=0$ and $\alpha=3.5$.}
	\label{ncq3}
\end{figure} 
However, for a quench to the blue dashed line of Fig. \ref{pd}, the scaling behavior may be different from $t^{-3/4}$. For example, for a quench from $(h,\chi)$=(3,1) to (0.5,0.90041), the scaling is
\begin{eqnarray}
	|\delta_{mn}(t)|\sim t^{-1.23},
\end{eqnarray}
  this is shown in Fig. \ref{ncq3}.
The reason for this result is also owing to the asymptotic behaviors of the energy spectrum and the fact $F_k^{(1)}$, which satisfy
\begin{eqnarray}
	&&\varepsilon_k-\varepsilon_{k_0}\propto(k-k_0)^{2.45}, \label{blue1}\\
	&&F_k^{(1)}\propto (k-k_0)^2,\label{blue2}
\end{eqnarray}
where $k_0=0$, as shown in Fig. \ref{ncq3}(b). This gives $p=2.45$ and $q=2$ 
for the integral (\ref{int3}), subsequently the scaling is $t^{-1.23}$. 

We also test  similar cases with different $\alpha$, the postquench  Hamiltonian is $(h,\chi)=(3,1)$, the parameters of the  postquench Hamiltonians and the  resulting scaling laws are 
\begin{eqnarray}
&(0.5,0.78486), \alpha=4.0,    & |\delta_{mn}(t)|\sim t^{-1.005},\\
&(0.5, 0.74649), \alpha=4.5, 	&|\delta_{mn}(t)|\sim t^{-0.834},\\
&(0.5,0.72940), \alpha=5.0, 	&|\delta_{mn}(t)|\sim t^{-0.765}, \\
&(0.5,0.71556), \alpha=6.0, 	&|\delta_{mn}(t)|\sim t^{-0.756}.
\end{eqnarray} 
It is obvious that the decaying exponent is close to $-3/4$ for large $\alpha$, but close to $-3/2$ for small $\alpha$.
The explanation for these results still boils down to the changes of the asymptotic behaviors of the energy spectrum.

\begin{figure}[htpb]
	\centering
	\includegraphics[width=0.75\columnwidth]{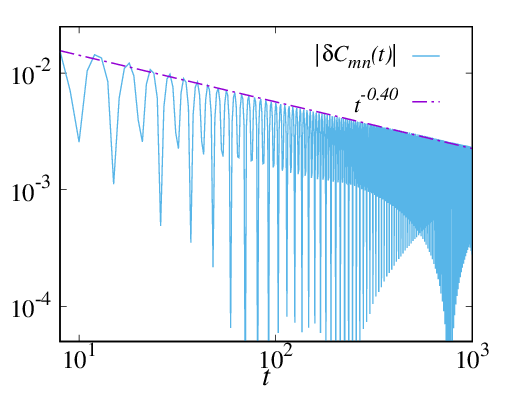}
	\caption{Scaling behavior of a quench from $(h,\chi)$=(3,1) to  (-1.5, 1.8076),  where the postqueched Hamiltonian is on the green dashed line of Fig. \ref{pd}.}
	\label{ncq4}
\end{figure} 
For a quench  to the green dashed line of Fig. \ref{pd},
the scaling of $|\delta C_{mn}(t)|$ is also different from $t^{-3/4}$.
For example, for a quench from $(h,\chi)$=(3,1) to (-1.5, 1.8076), the scaling is 
\begin{eqnarray}
 |\delta_{mn}(t)|\sim t^{-0.40},
\end{eqnarray}
 as shown in Fig. \ref{ncq4}.  
It is very important to note that in the current case, the result  does not fall in the prediction of the SPA.
In  the energy spectrum of the postquench Hamiltonian, there are three points  that satisfy the condition $d\varepsilon_k/dk=0$, they are $k=0$, $k=\pi$,
and $k=k_i=0.26625$, as shown in Fig. \ref{spec0}(b). Here $k=0$ and $k=\pi$ are saddle points, and $k=k_i$ is an inflection point.
In the vicinity of the saddle point $k=0$, we find that $\varepsilon_k \sim k^{1.65}$ and $F_k^{(1)} \propto k^2$, i.e., $p\approx1.65$ and $q=2$
 for the integral (\ref{int3}), which should give a $t^{-1.82}$ scaling. Similar analysis can be performed for the saddling point $k=\pi$, which should give a $t^{-3/2}$ scaling.  
We can see that both the scaling from $k=0$ and the scaling from $k=\pi$ are obviously different from the scaling $t^{-0.40}$. 
For the inflection point $k_i$, we find that in the vicinity of this point $\varepsilon_k-\varepsilon_{k_i} \propto (k-k_i)^3$ and $F_k^{(1)}\sim {\rm constant}$, i.e., $p=3$ and  $q=0$; however, in  this case we should not apply these values of $p$ and $q$ for the integral (\ref{int3}) to get a $t^{-1/3}$ scaling, because the SPA is not proved to be valid for such an inflection point. 


 We further test several other cases on such green dashed line, with different value of $\alpha$,  the postquench  Hamiltonian is $(h,\chi)=(3,1)$, 
 the parameters of the  postquench Hamiltonians and the  resulting scaling laws are 
 \begin{eqnarray}
 	&(-1.5, 1.7227), \alpha=4.0,    & |\delta_{mn}(t)|\sim t^{-0.474},\\
 	&(-1.5, 1.6663), \alpha=4.5, 	&|\delta_{mn}(t)|\sim t^{-0.594},\\
 	&(-1.5, 1.6310), \alpha=5.0, 	&|\delta_{mn}(t)|\sim t^{-0.689},\\
 	&(-1.5, 1.6001), \alpha=6.0, 	&|\delta_{mn}(t)|\sim t^{-0.737}.
 \end{eqnarray} 
All the decaying exponents are different from -3/4, especially in the cases where $\alpha$ is relatively small.

 In summary, the quench to the green dashed line is affected by an inflection point of the spectrum, which leads to different scaling law that is out of the prediction of the theory of SPA.

\section{Critical quench}
\label{cri}
\subsection{Quench to a critical point}
For a quench to the critical point, the long-range XY model shows different behaviors for the quench to the critical line 1 and the critical line  2 (as shown in Fig. \ref{pd}).
For the quench from a noncritical point to the critical line 1, the two typical examples we studied for $\alpha=3.5$ are 
\begin{eqnarray}
&(h,\chi)=(3, 1)\rightarrow (1, 1), &  |\delta_{mn}(t)|\sim t^{-3/2},\\
&(h,\chi)=(3, 1)\rightarrow (1, 2), & |\delta_{mn}(t)|\sim t^{-1/2},
\end{eqnarray}
we can see that the scaling laws are the same as the counterparts of the noncritical quench, and also the same as the counterparts of the short-range XY model. 
We also test other cases, we find  that the scaling laws are independent of the value of $\alpha$ (we only concern the case of $\alpha>1$).

\begin{figure}[tpb]
	\centering
	\includegraphics[width=1\columnwidth]{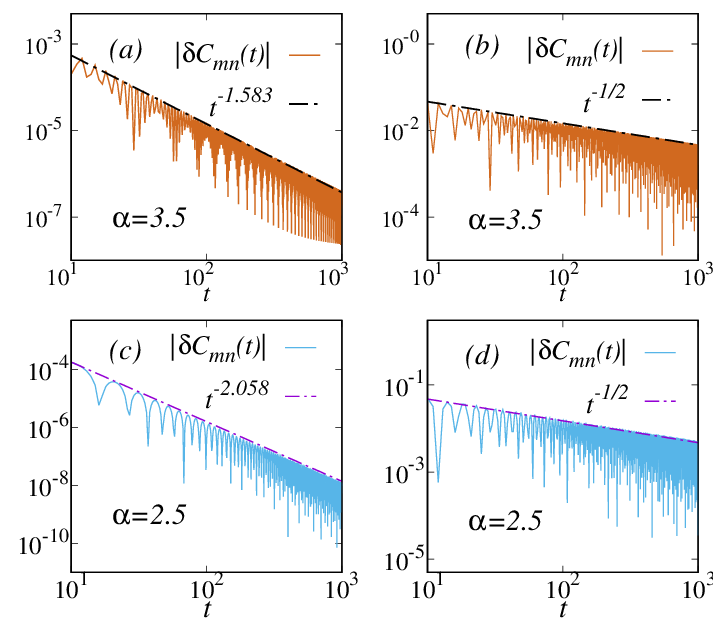}
	\caption {(a) and (b): quench from $(h,\chi)$=(3,1) to ($h_{\rm c}^{(2)}$, 1) and ($h_{\rm c}^{(2)}$,  2), respectively,  with $\alpha=3.5$ and $h_{\rm c}^{(2)}=-0.82322$;
		(c) and (d): quench from $(h,\chi)$=(3, 1) to ($h_{\rm c}^{(2)}$, 1) and ($h_{\rm c}^{(2)}$,  2), respectively,  with $\alpha=2.5$ and $h_{\rm c}^{(2)}=-0.64645$.}
	\label{cq2}
\end{figure}

We then investigate the quench to the critical line 2, in this case, we find that the scaling law of $|\delta C_{mn}(t)|$ may be changed.
For example, Fig. \ref{cq2}(a) shows a quench from (3, 1) to $(h_{\rm c}^{(2)}, 1)$, with $\alpha=3.5$,  in which the postquench Hamiltonian is a commensurate phase.
We can see that in this case the scaling law 
\begin{eqnarray}
	|\delta_{mn}(t)|\sim t^{-1.583}
\end{eqnarray}
is different from the noncritical quench shown in Fig. \ref{ncq1a}(a).
The deviation of the scaling  from $t^{-3/2}$ becomes more obvious when $\alpha$ becomes smaller,  for example, when $\alpha=2.5$, the scaling becomes 
\begin{eqnarray}
	|\delta_{mn}(t)|\sim t^{-2.058},
\end{eqnarray}
 as shown in Fig. \ref{cq2}(c). The results for these results can also be attributed to the asymptotic behaviors of the energy spectrum and the factor $F_k^{(1)}$ of the postquench Hamiltonian.  For example, in  the case  of Fig. \ref{cq2}(c), we find that 
 \begin{eqnarray}
 &&\varepsilon_k-\varepsilon_0\propto (k-0)^{1.42},\\
 &&F_k^{(1)}\propto (k-0)^{1.92},
 \end{eqnarray}
this gives $p=1.42$ and $q=1.92$ for the integral (\ref{int3}), which leads to the  $ t^{-2.058}$ scaling. It should be noted that in the current case, 
we consider the contribution of the saddle point $k=0$, the gap-closing point $k=\pi$ is not a saddle point, although it is also a local extreme point.

The long-range interactions only shelter the asymptotic behavior of the energy spectrum near the saddle point $k=0$ but not affect the point $k=\pi$,
this is the main reason for the difference in the results of the quench to the two critical lines; for the quench to the critical line 1, because the gap-closing point is $k=0$, 
thus the dominating saddle point is $k=\pi$, subsequently the scaling law of relaxation behavior is kept; 
on the contrary, for the quench to the critical line 2, the gap-closing point is $k=\pi$, and the dominant saddle point is $k=0$, therefore, the long-range effect in the energy spectrum manifests itself in the scaling behaviors.

At last, we show that for a quench to a point on the critical line 2 that is an incommensurate phase, because of the existence of an extra saddle point $k_0$
which is not equal to 0 or $\pi$, the scaling $t^{-1/2}$ does not change. Two typical examples are given in Figs. \ref{cq2}(b) and (d).

\subsection{Quench from a critical point}
For a quench from a critical point, the scaling law may be changed.
For example, with $\alpha=3.5$, a quench from the critical point $(h,\chi)$=(1, 1) to a commensurate phase with  $(h,\chi)$=(1.5, 1),  gives a scaling of 
\begin{eqnarray}
	|\delta_{mn}(t)|\sim t^{-1}.
\end{eqnarray}
This can be compared to the noncritical quench shown in Fig. \ref{ncq1a}(a), where the scaling law is $t^{-3/2}$.  
The reason for this result is related to the gap-closing of the prequench Hamiltonian, in this case $\sin\zeta_k\sim 1$, thus $F_k^{(2)}\sim k$, and subsequently $q=1$; meanwhile, the energy spectrum of the postquench Hamiltonian satisfies $\varepsilon_k-\varepsilon_0\sim k^2$, thus $p=2$. Therefore, the integral (\ref{int3}) gives a $t^{-1}$ scaling.
However, for a quench from a critical state to the incommensurate phase, the scaling law does not change; for example, 
for a quench from  $(h,\chi)$=(1, 1)  to (0.5, 0.5), the scaling is the same as the counterpart of the noncritical quench, i.e.,  $|\delta C_{mn}(t)|\sim t^{-1/2}$ ;
the reason for this result can be analyzed by the similar way.   The scaling law in this case is also not affected by the change of $\alpha$.

We then study the quench from a critical point to the boundary between the commensurate and incommensurate phases,
in this case, the scaling law may be changed or not, depending on the choosing of the parameters of both the prequench and postquench Hamiltonians.
For example, for the quench to the red dashed line, the two cases we studied are
\begin{eqnarray}
(h,\chi)=(1,1)\rightarrow(0.5, 1.2316),                              |\delta_{mn}(t)|\sim t^{-3/4},\label{cb1}\\
(h,\chi)=(h_{\rm c}^{(2)},1)\rightarrow(0.5, 1.2316),  |\delta_{mn}(t)|\sim t^{-1/2},\label{cb2}
\end{eqnarray}
where $\alpha=3.5$ and $h_{\rm c}^{(2)}=-0.82322$.  We can see that the scaling exponent of the quench  in Eq. (\ref{cb1}) is the same as the counterpart of the noncritical quench (as shown in Eq. \ref{ncred})); the reason is that the saddle point dominating the quench to this line is $k=\pi$, while the gap-closing point of prequench Hamiltonian is $k=0$, therefore the gap-closing property of the prequench Hamiltonian does not manifests itself in the scaling behavior. However, for the quench in Eq. \ref{cb2}, the gap-closing point is $k=\pi$,  which is the same as the dominant saddle point, it leads to the fact that $F_k^{(2)}\sim k-\pi$, when combined with the fact that $\varepsilon_k-\varepsilon_0\sim (k-\pi)^2$, according to the theory of SPA, the scaling should be $t^{-1/2}$.
 The scaling laws in Eqs. (\ref{cb1}) and (\ref{cb2})  are also the same as the counterparts of the short-range XY model\cite{Ding2023}, and remain unchanged when  $\alpha$ changes.

We then study the quench from the critical lines to the blue dashed line, the two cases we studied are 
\begin{eqnarray}
	(h,\chi)=(1,1)\rightarrow(0.5, 0.90041),                              |\delta_{mn}(t)|\sim t^{-0.822},\label{c2b1}\\
	(h,\chi)=(h_{\rm c}^{(2)},1)\rightarrow(0.5, 0.90041),  |\delta_{mn}(t)|\sim t^{-1.22},\label{c2b2}
\end{eqnarray}
where $\alpha=3.5$ and $h_{\rm c}^{(2)}=-0.82322$.
Comparing to the noncritical quench in Fig. \ref{ncq3}(a), the quench of (\ref{c2b1}) satisfies a different scaling. 
The reason is that in this case, the dominant saddle point of the postquench Hamiltonian is $k=0$, and the gap-closing point of the prequench Hamiltonian is
also $k=0$, this  leads to the change of the asymptotic behavior of $F^{(2)}_k$ from $F^{(2)}_k\propto k^2$ to $F^{(2)}_k\propto k$, combining with the fact that
$\varepsilon_k-\varepsilon_0\sim (k-0)^{2.45}$, we get the scaling  $t^{-0.822}$.
However, in the quench of (\ref{c2b2}), we can see that the scaling is the same as that in Fig. \ref{ncq3}(a), the reason is that here  the gap-closing point of the prequench Hamiltonian is $k=\pi$, which is different from the dominant saddle point $k=0$, 
it does not change the asymptotic behaviors of the energy spectrum of the postquench Hamiltonian and the factors  $F^{(1)}_k$ and $F^{(2)}_k$, therefore the scaling law keeps unchanged.

At last, we study the critical quench to the green dashed line,  in this case, we also find new scaling laws;  we studied two typical cases, the first case is
\begin{eqnarray}
	(h,\chi)=(1,1)\rightarrow(-1.5, 1.8076),                           |\delta_{mn}(t)|\sim t^{-0.339},\label{c2r1}
\end{eqnarray}
where $\alpha=3.5$. In this case, the energy spectrum of the postquench Hamiltonian has an inflection point, as shown in Fig. \ref{spec0}(b), which makes the resulting scaling law out of the prediction of the theory of SPA. We can also see that the scaling exponent is also different from the one of noncritical quench shown in Fig. \ref{ncq4},
this should be attributed to the gap-closing property of the postquench Hamiltonian.

\begin{figure}[htpb]
	\centering
	\includegraphics[width=0.75\columnwidth]{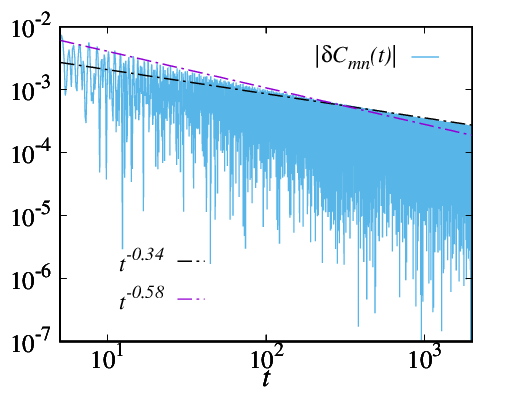}
	\caption{Relaxation behavior of a quench from $(h,\chi)=(h_{\rm c}^{(2)},1)$ to  (-1.5, 1.8076) with $\alpha=3.5$,  where the postqueched Hamiltonian is on the green dashed line of Fig. \ref{pd} and $h_{\rm c}^{(2)}=-0.82322$.}
	\label{hunhe}
\end{figure} 
 The second case we  studied is
\begin{eqnarray}
&&(h,\chi)=(h_{\rm c}^{(2)},1)\rightarrow(-1.5, 1.8076), \nonumber\\
&&|\delta_{mn}(t)|\sim at^{-0.34}+bt^{-0.58},\label{c2r2}
\end{eqnarray}
where $\alpha=3.5$ and $h_{\rm c}^{(2)}=-0.82322$; $a$ and $b$ are nonuniversal constants.  In this case, we can see that the relaxation behavior is a mixture of the $t^{-0.34}$ scaling and $t^{-0.58}$ scaling,  
where $t^{-0.58}$ dominates the behavior of the long-time limit and $t^{-0.34}$  dominates the behavior of the earlier time; 
this is demonstrated in Fig. \ref{hunhe}.  Such crossover behavior is obtained through  numerical calculations of Eqs. (\ref{int1}) and (\ref{int2}) and the data fitting. Currently, we only know that it is related to the emergence of the inflection point in the postquench Hamiltonian and the gap-closing property of the prequench Hamiltonian, but there is a lack of systematic theoretical explanation.

\section{Summary and discussion}
\label{con}
In summary, we have studied the real-time relaxation behavior of a long-range XY chain following a quantum quench. 
Comparing to the short-range model,  the long-range interactions can lead to some interesting changes  in the universal relaxation behaviors, 
 which are related to the  different structures of the energy spectrum. Especially, the long-range interactions can lead to inflection point in the energy spectrum, which makes  the relaxation behavior go beyond the prediction of the  stationary phase approximation. Our researches also include the critical quench,  the gap-closing  property of the initial state and the special energy spectrum structure of the postquench Hamiltonian may lead to different relaxation behaviors.

It is an interesting question to generalize the research of the current paper to other long-range models, such as the long-range Kitaev chain\cite{LRKC}, the Lipkin-Meshkov-Glick model\cite{LMG}, and also nonintegrable long-range models\cite{longIsing,longHeisenberg}.  Another related interesting question is the generalization to quasiperiodic and disordered models\cite{quasiIsing,disorderIsing}, which may have complicated structures of energy spectrum; because the relaxation behavior is mainly determined by the structure of the energy spectrum,  it is expected that the relaxation behaviors in these models may be very different.

\section*{Acknowledgment}
This work is supported by the National Natural Science Foundation of China under Grant No. 11975024 and the Anhui Provincial Supporting Program for Excellent Young Talents in Colleges and Universities under Grant No. gxyqZD2019023.

\end{document}